\documentclass[sigconf]{acmart}

\usepackage{amsmath,amssymb,amsfonts}
\usepackage{graphicx}
\usepackage{textcomp}

\usepackage{booktabs}
\usepackage{multirow}
\usepackage{algorithm}
\usepackage[noend]{algpseudocode}

\usepackage{blindtext}
\usepackage{graphicx}
\usepackage{wrapfig}
\usepackage[caption=false]{subfig}
\usepackage[font=footnotesize]{caption}

\usepackage{tabularx}
\usepackage{amssymb}
\usepackage{soul}

\usepackage{enumitem}

\usepackage{pifont}

\usepackage[normalem]{ulem}

\usepackage{lipsum}  
\usepackage{diagbox}

\renewcommand\footnotetextcopyrightpermission[1]{} % removes footnote with conference information in first column

\AtBeginDocument{%
  \providecommand\BibTeX{{%
    \normalfont B\kern-0.5em{\scshape i\kern-0.25em b}\kern-0.8em\TeX}}}

\copyrightyear{2020}
\acmYear{2020}
\setcopyright{acmcopyright}\acmConference[ICCAD '20]{IEEE/ACM International Conference on Computer-Aided Design}{November 2--5, 2020}{Virtual Event, USA}
\acmBooktitle{IEEE/ACM International Conference on Computer-Aided Design (ICCAD '20), November 2--5, 2020, Virtual Event, USA}
\acmPrice{15.00}
\acmDOI{10.1145/3400302.3415667}
\acmISBN{978-1-4503-8026-3/20/11}

\begin{document}

\title{InterLock: An Intercorrelated Logic and Routing Locking}


% \author{Hadi M Kamali, Kimia Z Azar}
% \affiliation{%
%   \institution{George Mason University}
%   \city{Fairfax, VA}
% \email{{hmardani, kzamiria}@gmu.edu}

% \author{Houman Homayoun}
% \affiliation{%
%   \institution{University of California Davis}
%   \city{Davis, CA}
% \email{hhomayoun@ucdavis.edu}

% \author{Avesta Sasan}
% \affiliation{%
%   \institution{George Mason University}
%   \city{Fairfax, VA}
% \email{asasan@gmu.edu}

\author{Hadi Mardani Kamali$^1$, Kimia Zamiri Azar$^1$, Houman Homayoun$^2$, Avesta Sasan$^1$}
\affiliation{\vspace{0.1cm}
\institution{$^1$ George Mason University, Fairfax, VA, USA.}
}
\email{{hmardani, kzamiria, asasan}@gmu.edu}
\affiliation{\vspace{0.1cm}
\institution{$^2$ University of California, Davis, Davis, CA, USA.}
}
\email{{hhomayoun}@ucdavis.edu}
\renewcommand{\shortauthors}{}
\renewcommand{\shorttitle}{}

\begin{abstract}

In this paper, we propose a \emph{canonical prune-and-SAT} (\emph{CP\&SAT}) attack for breaking state-of-the-art routing-based obfuscation techniques. In the \emph{CP\&SAT} attack, we first encode the key-programmable routing blocks (keyRBs) based on an efficient SAT encoding mechanism suited for detailed routing constraints, and then efficiently re-encode and reduce the CNF corresponded to the keyRB using a bounded variable addition (BVA) algorithm. In the \emph{CP\&SAT} attack, this is done before subjecting the circuit to the SAT attack. We illustrate that this encoding and BVA-based pre-processing significantly reduces the size of the CNF corresponded to the routing-based obfuscated circuit, in the result of which we observe 100\% success rate for breaking prior art routing-based obfuscation techniques. Further, we propose a new \emph{intercorrelated logic and routing locking} technique, or in short \emph{InterLock}, as a countermeasure to mitigate the CP\&SAT attack. In Interlock, in addition to hiding the connectivity, a part of the logic (gates) in the selected timing paths are also implemented in the keyRB(s). We illustrate that when the logic gates are twisted with keyRBs, the BVA could not provide any advantage as a pre-processing step. Our experimental results show that, by using InterLock, with only three 8$\times$8 or only two 16$\times$16 keyRBs (twisted with actual logic gates), the resilience against existing attacks as well as our new proposed \emph{CP\&SAT} attack would be guaranteed while, on average, the delay/area overhead is less than 10\% for even medium-size benchmark circuits. 

\end{abstract}


\settopmatter{printfolios=true}

\maketitle

\pagestyle{plain}

% {\fontsize{8pt}{8pt} \selectfont
% \textbf{ACM Reference Format:} \\
% Hadi M Kamali, Kimia Z Azar, Houman Homayoun, and Avesta Sasan.
% 2020. InterLock: An Intercorrelated Logic and Routing Locking. In \it{IEEE/ACM International Conference on Computer-Aided Design (ICCAD ’20), November 2–5, 2020, Virtual Event, USA.} ACM, New York, NY, USA, 9 pages. \\ https://doi.org/10.1145/3400302.3415667 }

\section{Introduction}

The globalization of the design and implementation of integrated circuits has drastically increased, particularly in the past two decades. This is when high-tech companies try (1) to reduce the cost of manufacturing, (2) to access technology that is inclusively available by a limited number of suppliers, (3) to reduce time to market, and (4) to meet the market demand \cite{yeh2012trends}. However, it has also raised many security threats and trust challenges. Some of these threats include that of IC overproduction, Hardware Trojan insertion, reverse engineering, and Intellectual Property (IP) theft \cite{rostami2014primer}.   

To combat these threats, numerous \emph{Design-for-Trust} (\emph{DfTr}) techniques have been proposed, one of them is \emph{Logic locking} \cite{roy2010ending, rajendran2012security}, \emph{a.k.a} logic obfuscation. In Logic locking, the designer adds post-manufacturing programmability into the design controlled by programmable values referred to as the \emph{key}. The key value is driven from an on-chip tamper-proof non-volatile memory (tpNVM) \cite{tuyls2006read}, and it will be initiated after fabrication via a trusted party. Hence, the adversary cannot recover the correct functionality of a logic locked chip without having the correct key.

The security and the strength of the primitive logic locking techniques \cite{roy2010ending, rajendran2012security, rajendran2015fault} has been called into question by various attacks, especially the \emph{Boolean satisfiability} (SAT) based attack \cite{subramanyan2015evaluating, el2015integrated}. In the SAT attack, it is assumed an adversary has access to (1) an oracle (working chip), (2) a fully reverse engineered netlist, and (3) the scan chain access of oracle (writing/reading the content of internal registers at will.). Based on these assumptions, the SAT attack, as an \emph{oracle-guided} attack, starts iteratively ruling out the set of incorrect keys using a few selected input queries found by the SAT solver, called discriminating inputs (DIPs) \cite{subramanyan2015evaluating, el2015integrated}.

To thwart the SAT attack, over the past few years, researchers have investigated four different categories \cite{zamiri2019threats}:

(1) \textbf{Point Function Based Obfuscation:} The first group of techniques, examples of which include SARLock, Anti-SAT, and SFLL \cite{yasin2016sarlock, xie2016mitigating, yasin2017provably}, tries to reduce the strength of the SAT attack such that each DIP could only rule out one incorrect key (or a few). So, it significantly increases the number of required SAT iterations (required DIPs). However, they are vulnerable against structural-based attacks \cite{yasin2017security, yasin2017removal, xu2017novel, sirone2020functional}. Besides, these techniques suffer from very low output corruption,  making them susceptible to approximate-based attacks \cite{shamsi2017appsat, shen2017double}. 

(2) \textbf{Behavioral/Cyclic Obfuscation:} In the second group of techniques, such as delay locking \cite{xie2017delay}, timing-based locking \cite{zhang2018timingcamouflage, alam2019toic}, or cyclic locking \cite{shamsi2017cyclic, rezaei2018cyclic, roshanisefat2018srclock, rezaei2019cycsat, roshanisefat2020sat}, the obfuscated circuit (I) is not translatable to a SAT problem (delay/timing based), or (II) traps the SAT solver in an infinite loop (cyclic), or (III) it leads to an incorrect key (cyclic). However, the existing techniques in this category are already broken using SMT attack (delay) \cite{azar2019smt}\nocite{azar2020nngsat}, timingSAT attack (timing) \cite{chakraborty2018timingsat}, and the SAT-based attacks on cyclic locking \cite{zhou2017cycsat, shen2019besat, shamsi2019icysat}. 

(3) \textbf{Scan Chain Blocking/Obfuscation:} Since many of prevailing attacks rely on access to the scan chain, the third group of techniques locks/blocks the scan for any unauthorized scan access \cite{wang2017secure, karmakar2018encrypt, limaye2019robust, roshanisefat2020dfssd, kamali2020scramble, kamali2020designing, azar2019coma}. Since the SAT attack is only applicable to combinational circuits, when the scan chain is blocked/obfuscated, the adversary's access is limited to only primary inputs/outputs (PI/PO). Hence, the SAT attack is no longer applicable to the whole (sequential) circuit. However, these techniques are later broken using unrolling-based SAT attacks as well as the SAT attack integrated with BMC \cite{el2017reverse, shamsi2019kc2, alrahis2019scansat}. Also, blocking the scan chain enforces the tester to rely on the PO for any test/debug purpose, which might reduce the test coverage considerably. 

(4) \textbf{Symmetric Interconnection Obfuscation:} Taking a step further, the fourth group of techniques tries to significantly increase the complexity of inner calculations of the SAT solver leading to an extremely long runtime per \emph{each iteration} of the SAT attack \cite{kamali2019full, shamsi2018cross, patnaik2020obfuscating}\nocite{kamali2018lut}. These techniques rely on building symmetric interconnection into the locked portion of the circuit, extremely increasing the depth of the SAT search tree, and reducing the number of derived variables (implications) based on assigned variables. The building block of the existing solutions in this category are the key-programmable routing blocks (we call them \emph{keyRB} in this paper), each has its topology, such as crossbar or permutation (logarithmic) network. Although this group of obfuscation techniques suffers from the higher area/delay overhead, to the best of our knowledge, there is still no attack on this category of logic locking techniques. 

\vspace{-5pt}
\subsection{Contribution}

In this paper, we first propose a \emph{canonical prune-and-SAT} (\emph{CP\&SAT}) attack on the fourth group of techniques (The first attack on routing-based obfuscation). In our proposed \emph{CP\&SAT} attack, we first extract and model the circuit into some numerical bound problems, which could be re-encoded efficiently using a bounded variable addition (BVA) algorithm. Then, the BVA algorithm is applied to each numerical bound problem, separately. The BVA re-encodes and reduces the CNF size/complexity of each numerical bound problem significantly. After reduction using the BVA, the reduced CNFs (corresponded to numerical bound problems) will be merged again with the circuit's CNF, and the SAT solver could be executed on the reduced CNF version. Our security analysis on benchmark circuits protected by existing routing obfuscation techniques, such as Cross-Lock and Full-Lock \cite{shamsi2018cross, kamali2019full}, demonstrates 100\% successfulness of this attack within a short time.

We then propose an enhanced obfuscated technique, that mitigates the weakness of existing routing-based obfuscation techniques \cite{shamsi2018cross, kamali2019full} against the proposed \emph{CP\&SAT} attack. We refer to our proposed obfuscation solutions as \emph{InterLock}. In \emph{InterLock}, the routing obfuscation (keyRB) is \emph{intercorrelated} with logic obfuscation. Hence, since the logic is truly twisted with routing all controlled by the key, it is not possible to convert and model the keyRB(s) into the numerical bound problem(s), and consequently,  the BVA is no longer applicable to them for reduction.

We implement and evaluate the keyRBs in \emph{InterLock} based on three different technologies: (1) transmission-gate (Tgate) CMOS, (2) programmable-via using anti-fuse elements (PVIA), and (3) three-independent-gate field-effect transistors (TIGFET). It helps us to provide a better illustration of the area/delay overhead. We also show that by implementing in the lower level of abstraction, the area/delay overhead of \emph{InterLock} could be even below $\sim$10\% to make the design resilient against the prevailing attacks.

% Commented for having more space.
% The remainder of the paper is organized as follows: In Section \ref{background}, we first describe how routing-based obfuscation works. Then, in Section \ref{proposed_attack}, we demonstrate how the \emph{CP\&SAT} splits and re-encodes the larger SAT problems into multiple smaller and easy to solve problems using the BVA mechanism. In Section \ref{proposed_defense}, we introduce \emph{InterLock}, as the new countermeasure against \emph{CP\&SAT} attack. We show how \emph{InterLock} can still get the benefit of routing-based obfuscation while it is also resilient against the \emph{CP\&SAT}. We elaborate our experiments in Section \ref{experiment}, and finally, the paper is concluded in Section \ref{conclusion}.  

\section{Background} \label{background}

% In the security evaluation of logic locking techniques, there are three basic assumptions: (1) The adversary has access to the reverse-engineered netlist of the locked circuit; (2) The adversary has access to an activated/unlocked chip; And (3) in the activated/unlocked chip, the scan chain access is open for test/debug purposes. These three assumptions are the substructure of the emergence of the Boolean satisfiability (SAT) based attack, which could be categorized as an \emph{oracle-guided} attack. 

\subsection{The Oracle-guided SAT Attacks}

In the SAT attack, with having access to (1) a working chip with open scan chain, and (2) the reverse-engineered netlist, each combinational part of the circuit could be evaluated independently. For any arbitrary combinational part (which is locked), ${c}_{lock}{: I\times K \rightarrow O}$, where $K = \{0,1\}^{k}$ is the \emph{key} space, there exists $k_{c} \in K$ such that $\forall i \in I\Rightarrow {c}_{lock}(i, k_c) = {c}_{oracle}(i)$, and ${c}_{oracle}$ is the  combinational logic part of working chip (\emph{oracle}). 

In the SAT attack, by getting inspiration from the miter circuit used in formal verification (equivalency checking), a miter circuit has been built as $miter \equiv {c}_{comb\_lock}(dip, k_1) \neq {c}_{comb\_lock}(dip, k_2)$. This miter circuit returns a specific \emph{discriminating input pattern} ($dip$) that produces different output using two different keys $k_1$ and $k_2$. Then, this $dip$ is queried on the oracle $eval \leftarrow {c}_{comb}(dip)$ and a new I/O constraint will be generated. This new I/O constraint ${c}_{comb\_lock}(dip, k_1) = {c}_{comb\_lock}(dip, k_2) = eval$ is stored back in the solver and the $miter$ circuit would be solved again to find a new $dip$. When the $miter$+constraints problem has no longer satisfying assignment (no new $dip$), the constraints could identify a $k_{c} \in K$.

\subsection{Point Function Based Obfuscation}

Based on the iterative structure of the SAT attack, its runtime could be obtained from:
\begin{equation} \label{runtime_simple}
T_{Attack}  = \sum_{i=1}^{N}T(i) = \sum_{i=1}^{N}(t_i + T_{DPLL}(\Phi_i)) 
\end{equation}
In Eq. \ref{runtime_simple}, the $T_{DPLL}$ is the runtime of the DPLL (\emph{Davis-Putnam-Logemann-Loveland}) algorithm, $t_i$ is the runtime of the remaining book-keeping code executed at each iteration $i$, and $N$ is the number of iterations (number of DIPs) required for de-obfuscation. The DPLL (or one of its derivatives) is a recursive algorithm that used to perform \emph{Conflict-Driven Clause Learning} (\emph{CDCL}) as the main part of the SAT solver. Getting the benefit of this recursive algorithm, the SAT attack can recover the correct functionality with fast convergence. 

The most intuitive mechanism to combat against the srength of the SAT attack is to find a way to maximize $N$, which is the main aim of point function based obfuscation techniques (category 1). \cite{yasin2016sarlock, xie2016mitigating, yasin2017provably}. For this purpose, the structures proposed in this category weakens the pruning power of each $dip$, guaranteeing that each $dip$ can only rule out one (or a small number of) incorrect key(s). This forces the number of needed iterations ($N$) to exponentially increase with respect to the number of keys. However, with a very shallow DPLL recursive tree, the execution time of each iteration of the SAT solver ($t + T_{DPLL}$) is quite short. 

\subsection{Symmetric Interconnection Obfuscation: One Step Deeper} 

Symmetric interconnection obfuscation techniques show how the direction of adding difficulties to the SAT attack could be changed (becoming deeper) via deepening the DPLL tree. This deepening could be achieved if the obfuscation circuit forces a relationship between the number of clauses and the number of variables to maximize the penalty associated with incorrect variable assignment symmetrically across the search tree. With having a deeper DPLL tree, the runtime formula of the SAT attack (Equation \ref{runtime_simple}) could be re-written as follows \cite{kamali2019full}: 
\begin{equation} \label{runtime_recursive}
T_{Attack}  =  \sum_{i=1}^{N}(t_i + T_{DPLL}(\Phi_i)) \simeq  \sum_{i=1}^{N}\sum_{j=1}^{M}(T_{DPLL}^{Avg})
\end{equation}
The main aim of symmetric interconnection obfuscation techniques (category 4) is to extremely increase the number ($M$) and the computational complexity ($T_{DPLL}^{Avg}$) of recursive calls in DPLL algorithm, which occurs when the DPLL tree is extremely deep/large enough. Two examples of this category are Cross-Lock \cite{shamsi2018cross} and Full-Lock \cite{kamali2019full}.

\subsubsection{Cross-Lock: PVIA-based Crossbar Obfuscation}
\begin{figure}[t]
    \centering
    \includegraphics[width = 245pt]{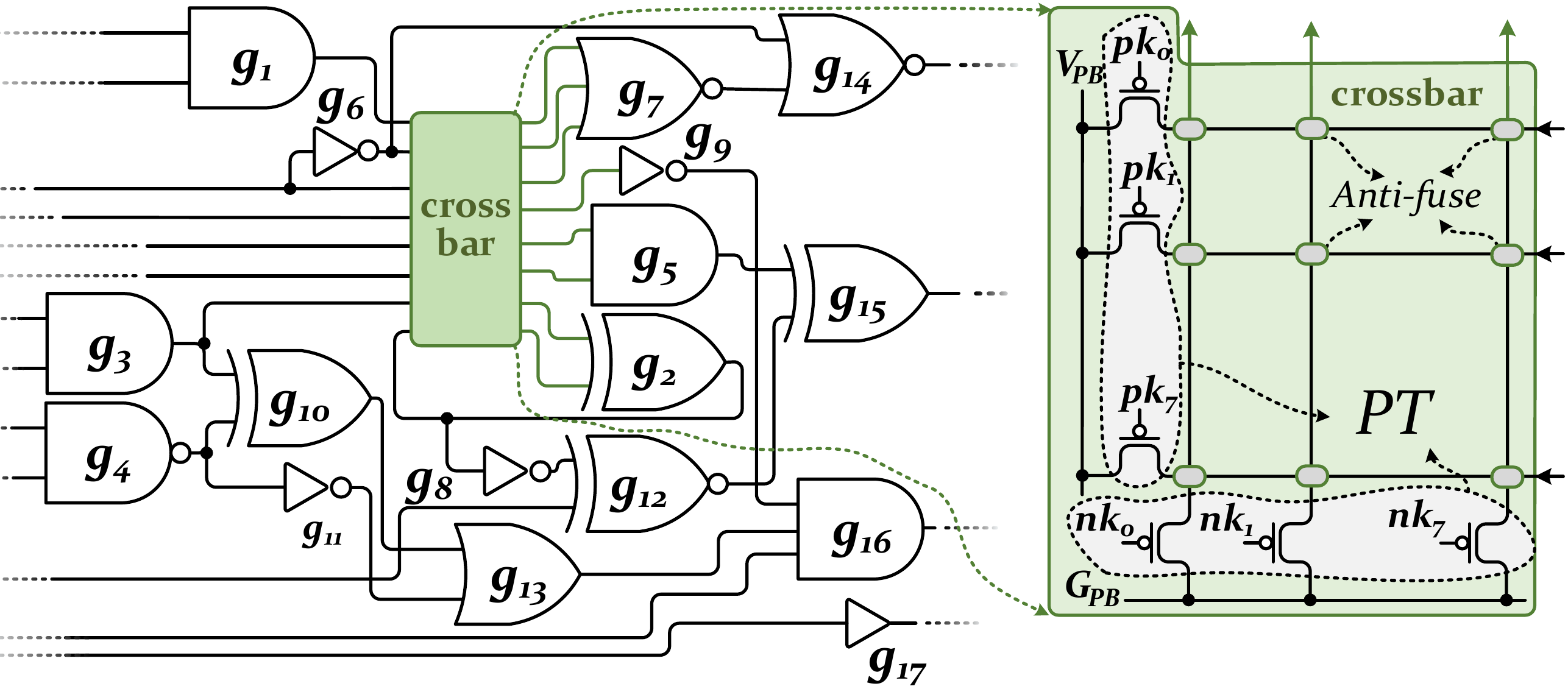}
    \vspace{-15pt}
    \caption{Circuit Locked by Cross-Lock \cite{shamsi2018cross} with an $8 \times 8$ Crossbar Network.}
    \vspace{-10pt}
    \label{cross_lock_example}
\end{figure}

\textcolor{white}{\vspace{8pt} \\}
In Cross-Lock \cite{shamsi2018cross}, each keyRB are built using programmable-vias (PVIA) constructed using one-time-programmable (OTP) elements made of anti-fuse \cite{actel2002design}. PIVAs are used to implement $n \times m$ crossbars. The main approach for PVIA programming is connecting two complementary (NMOS and PMOS) programming transistors (PTs) to the two ends that connect the device terminals to programming supplies \cite{actel2002design}. Routing obfuscation in Cross-lock has been performed by inserting PVIA-based $n \times m$ crossbars. Fig. \ref{cross_lock_example} shows a small circuit locked by Cross-Lock with $n=m=8$. 

\subsubsection{Full-Lock: Logarithmic-based Logic+Routing Obfuscation}
\begin{figure}
    \centering
    \includegraphics[width = 240pt]{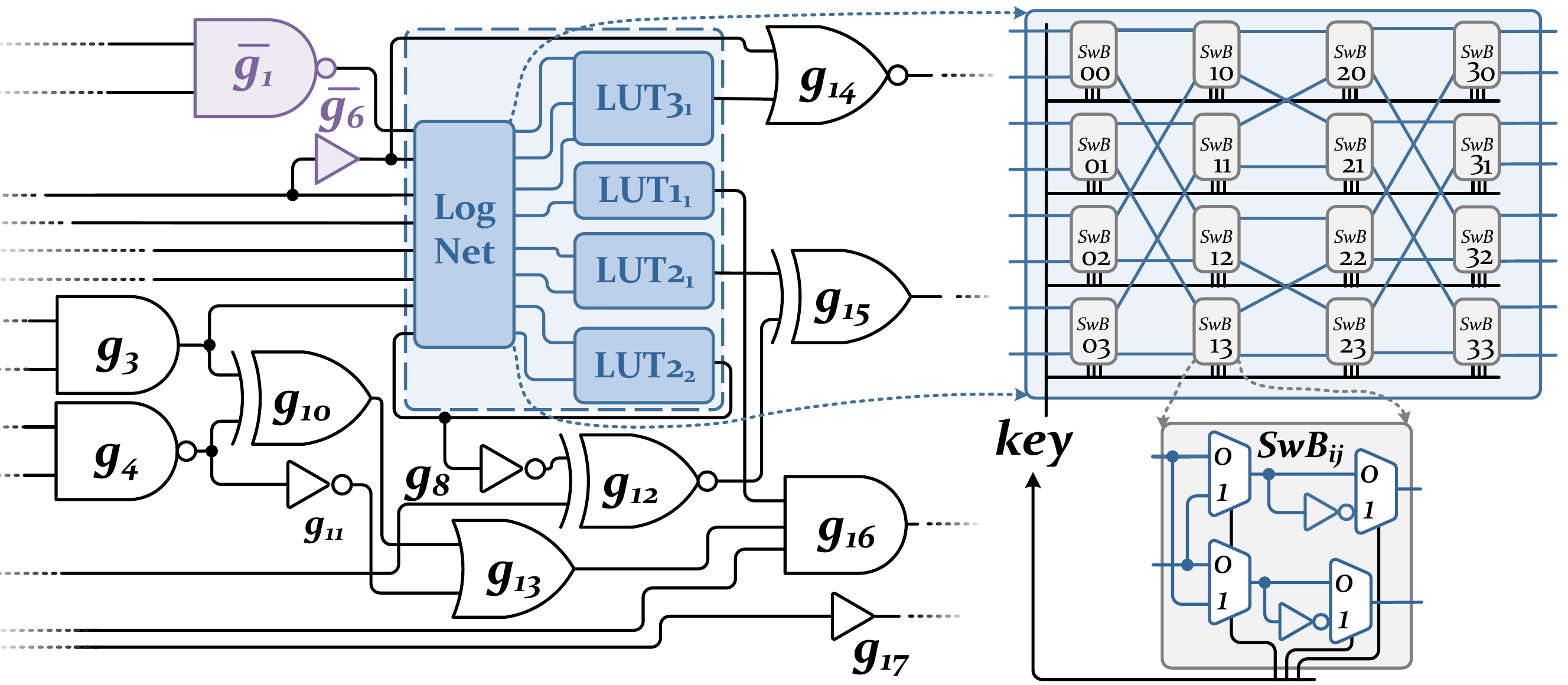}
    \vspace{-15pt}
    \caption{Circuit Locked by Full-Lock \cite{kamali2019full} with an $8 \times 8$ Logarithmic Network.}
    \vspace{-10pt}
    \label{full_lock_example}
\end{figure}

\textcolor{white}{\vspace{8pt} \\}
In Full-Lock \cite{kamali2019full}, the keyRB(s) is built using logarithmic near non-blocking routing network. Since the SAT solver accepts the inputs in conjunctive normal form (CNF), Full-Lock was motivated by \cite{mitchell1992hard}, where it was shown that the number of recursive calls in the DPLL algorithm could be maximized when the \emph{clause to variable ratio} of a CNF is close to 4. Then, a symmetric logarithmic routing network is introduced in Full-Lock that forces this ratio to 4. Also, to elevate the complexity of the obfuscated circuit against the SAT attack, the logic gates succeeding each keyRB are replaced with same-size look-up-tables (LUT). Fig. \ref{full_lock_example} shows a small circuit (similar circuit with Fig. \ref{cross_lock_example}) that is locked by Full-Lock with $n=m=8$, and the succeeding gates ($g_2$, $g_5$, $g_7$, and $g_{9}$) are replaced with same-size LUTs ($LUT_{2(1)}$, $LUT_{2(2)}$, $LUT_{3(1)}$, and $LUT_{1(1)}$, respectively). Also, in each key-programmable switch-box ($SwB$), a key-programmable inverter has been added, allowing output to be negated based on the key value. This negation capability allows us to replace the logic gates preceding each keyRB with their negated version ($g_1$: \emph{AND} $\rightarrow$ \emph{NAND}, $g_6$: \emph{NOT} $\rightarrow$ \emph{BUFF}), and handle the negation within the keyRB. 

\subsection{Cases Requiring Cyclic-based SAT Attack}

By using a keyRB for routing-based obfuscation, when the selected nets are correlated, an incorrect key might generate a feedback, which results in the generation of a combinational cycle. For example, when a net is in the fan-in-cone of another net, and both are selected as input to the keyRB, an incorrect key most likely generates a combinational cycle (e.g. $g_2$ $\rightarrow$ \emph{crossbar} $\rightarrow$ $g_2$ in Fig. \ref{cross_lock_example}). 

Since the SAT attack is only applicable to \emph{directed acyclic graphs} (DAG), the generation of cycles mislead (to an incorrect key, or an infinite loop) the SAT attack. The existence of cycles, however, does not prevent the SAT attack formulation. In many studies on cyclic obfuscation \cite{zhou2017cycsat, shamsi2019icysat, shen2019besat}, it was shown that by adding a pre-processing step to the SAT attack, it could add necessary cycle avoidance clauses for a successful SAT attack in the presence of combinational cycles. From this argument, for the security analysis of routing obfuscation, a cyclic-based SAT attack must be used.    

\section{Canonical Prune-and-SAT Attack} \label{proposed_attack}

As of today, there is still no successful attack on routing-based obfuscation. Each iteration of SAT solving on routing-based obfuscated circuits faces an ultra-deep and complex DPLL tree. Hence, the SAT attack cannot even find a satisfying assignment(s) to complete the de-obfuscation process. However, in this work, we propose \emph{canonical prune-and-SAT} (\emph{CP\&SAT}) attack on the routing-based obfuscation techniques. In the \emph{CP\&SAT}, we first model the key-programmable routing blocks (keyRB(s)) as numerical bound problems, and then a bounded variable addition (BVA) algorithm has been engaged as a pre-processing step to reduce the size and complexity of numerical bound problems. By using the BVA algorithm, the CNF corresponded to each keyRB will be reduced dramatically in terms of the number of clauses. Then, the re-encoded CNF will be solved using the traditional SAT attack.  

\subsection{Threat Model in \emph{CP\&SAT} Attack} \label{threat}

The \emph{CP\&SAT} attack will be performed based on the conventional threat model for logic locking \cite{subramanyan2015evaluating, el2015integrated, rajendran2012security}, where:

\begin{enumerate}[leftmargin=*]
    \item The adversary has access to the successfully reverse-engineered yet locked netlist. Hence, (s)he has all the necessary information about the netlist, such as the obfuscation technique, the key gates, the key inputs, etc. Specifically in routing obfuscation, the location of the keyRBs could be determined by the adversary. 
    \item The adversary has access to an activated/unlocked chip, in which the correct key is embedded into a secure tpNVM. 
    \item With having scan chain access on the activated/unlocked chip, the adversary can apply the SAT attack on each combinational part of the circuit, independently. 
\end{enumerate}

\subsection{Attack Flow}

The proposed \emph{CP\&SAT} attack is composed of three main steps: (1) modeling the keyRB(s) to be presented as a numerical bound problem, where for each output of keyRB, a sub-CNF will be extracted from the CNF of the whole circuit; (2) re-encoding the sub-CNF corresponded to each keyRB output using bounded variable addition (BVA) algorithm; (3) merging the updated (reduced) sub-CNFs into the CNF of the whole circuit, running the traditional SAT attack, and match the key for the correct routing.

\subsubsection{Modeling keyRB as a Numerical Bound Problem}
\textcolor{white}{\vspace{8pt} \\}
Extensive analysis on the application of Boolean satisfiability in detailed routing constraints \cite{aloul2002solving, chai2005fast, velev2008comparison, nam2004comparative, velev2009efficient} shows that the SAT solvers can consider simultaneously the routability constraints for all nets, leading to potentially faster convergence to a solution. However, this only happens when an appropriate \emph{encoding approach} has been chosen to represent routing constraints as a SAT problem before solving. Many studies have investigated and compared different encoding approaches \cite{velev2008comparison, nam2004comparative}. Using the key observations provided in these studies, in the first step of the proposed \emph{CP\&SAT} attack, we encode the sub-CNF related to each keyRB output using \emph{one-layer linear encoding}. To describe the logic-equivalent model, for each output of a $n \times n$ keyRB, the one-layer linear encoding replaces the original sub-CNF with a CNF describing a one-layer $n-to-1$ multiplexer (MUX) controlled by the one-hot key. More formally, for a $n \times n$ keyRB, the sub-CNF of each keyRB output, which is encoded using one-layer linear encoding, will be as follows:  
\begin{equation} \label{cardinality}
\bigwedge_{M \subseteq {1, ..., n} ,\atop {\mid M \mid = 1 }} {\big{(}}\bigvee_{i \in M} x_{i}k_{i}{\big{)}}
\end{equation}

In which $x_i$ denotes the wire that is connected to the $i^{th}$ input of the keyRB, and $k_i$ denotes the one-hot key that connects the $i^{th}$ input of the keyRB to the corresponded keyRB output when it is 1, and $M$ is the search space for each keyRB output. The Eq. \ref{cardinality} is the most special case of encoding of numerical bounds \cite{manthey2012automated}. The numerical bound problems could be denoted as  $\leq p(x_1, x_2, ..., x_n)$, meaning that among $n$ variables $p$ variables are allowed to be assigned true. The most special case of numerical bounds is when $p=1$, called \emph{at-most-1 constraint}, that is applied whenever a finite domain is encoded, and the Eq. \ref{cardinality} is one form of at-most-1 constraint encoding. According to this encoding definition, in the first step of the proposed \emph{CP\&SAT} attack, we first extract the sub-CNF related to each output of keyRB(s). Then, we use \emph{one-layer linear encoding} for the extracted sub-CNF to be encoded as a numerical bound problem. Then, in the second step as described in the next section, we use the BVA to re-encode and reduce the size of each sub-CNF for each output of the keyRB.

\subsubsection{SAT Reduction using Bounded Variable Addition}
\textcolor{white}{\vspace{8pt} \\}
As an integral part of SAT solving, \emph{resolution} and \emph{variable elimination} (VE) are two rules that would be applied on CNF before running the SAT solver to reduce the size of variables/literals \cite{davis1960computing, biere2004resolve, een2005effective}. The VE, as a proof procedure for CNF formulas, faces an exponential space complexity. Hence, to make it practical for usage, the VE must be bounded \cite{biere2004resolve, een2005effective}. In bounded VE (BVE) a variable $x$ could be eliminated only if $|S| \leq |S_{x} \cup S_{\bar{x}}|$, in which $S_{x} (S_{\bar{x}})$ denotes a set containing clauses all contain $x (\bar{x})$, $S$ is obtained from Eq. \ref{resolving}, and $|S| \leq |S_{x} \cup S_{\bar{x}}|$ means that the resulting CNF {\big{(}}$((F\setminus(S_{x} \cup S_{\bar{x}})) \cup S${\big{)}}\footnote{$((F\setminus(S_{x} \cup S_{\bar{x}})) \cup S$ means that, in CNF $F$, both sets of clauses, $S_{x}$ and $S_{\bar{x}}$, that contain $x$ and $\bar{x}$, respectively, must be replaced with clauses of set $S$ that are built using Eq. \ref{resolving}.} will contain no more than the original CNF ($F$) clauses.
\begin{equation} \label{resolving}
S  =  S_x \otimes S_{\bar{x}} = \{C_1 \otimes C_2 | C_1 \in S_x, C_2 \in S_{\bar{x}}, C_1 \otimes C_2 \neq Tautology\} 
\end{equation}

In \emph{CP\&SAT attack}, we engage the complementary version of BVE, called \emph{bounded variable addition} (BVA) \cite{manthey2012automated}, in which either a new variable will be added to the CNF or a variable will be substituted. Similar to BVE, the same \emph{bounding} concept must be used in BVA to decrease the size of the CNF \cite{manthey2012automated}. As the simplest example of the BVA, by adding a new variable $x$ to the following formula $F$ with 6 clauses, the re-encoded formula $F'$ would have one clause less.
\begin{equation} \label{sample_cnf}
F  = (a \lor c) \land (a \lor d) \land (a \lor e) \land (b \lor c) \land (b \lor d) \land (b \lor e)
\end{equation}
\begin{equation} \label{sample_reencoded_cnf}
F'  = (a \lor x) \land (b \lor x) \land (c \lor \bar{x}) \land (d \lor \bar{x}) \land (e \lor \bar{x})
\end{equation}
In the BVA, the number of possibilities to add or substitute a variable is extremely large. Hence, to make it practical for any CNF ($F$), the BVA algorithm must be constructed based on two steps: 

\begin{enumerate}[leftmargin=*]
    \item \emph{Replaceable Matching}: Creating a pair of sets consisting of a set of literals ($SET_{L}$) and a set of clauses ($SET_{C}$) such that for all $\{l, c\} \in  \{SET_{L}, SET_{C}\}$, the clauses $(c \setminus \{SET_{L}\}) \cup \{l\}$ are either in CNF ($F$) or tautological.
    \item \emph{matching-to-clauses}: Using a method that creates the sets $S_x = \{(l \lor x)~|~l \in SET_{L}\}$ and $S_{\bar{x}} = \{(c \setminus SET_{L}) \cup \{\bar{x}\}~|~c \in SET_{C}\}$, and removes all clauses $(c \setminus \{SET_{L}) \cup \{l\}$, and replaces them with $S_x \cup S_{\bar{x}}$.
\end{enumerate}

By applying these two steps on $F$ (Eq. \ref{sample_cnf}), $SET_{L} = \{a, b\}$ and $SET_{C} = \{(a \lor c), (a \lor d), (a \lor e)\}$, $F'$ could be generated using \emph{matching-to-clauses} with one clause less as shown in Eq. \ref{sample_reencoded_cnf}. Now, by using this 2-step BVA algorithm, any CNF formula could be reduced provably in size (the number) of clauses while the reduced CNF is also provably equivalent with the original CNF. 

\textbf{Theorem 1.} \emph{For two sets as \emph{replaceable matching} $\{SET_{L}, SET_{C}\}$ as for the CNF formula $F$, $F'$ as the reduced of $F$ could be constructed by adding a Boolean variable such that (1) $F'$ is logically equivalent to $F$ and (2) $F'$ contains $|F'| + |SET_{C}|+|SET_{L}| - |SET_{C}|\times|SET_{L}|$ clauses if none of the resolvents is a tautology.}

\emph{Proof.} For two sets as replaceable matching $\{SET_{L}, SET_{C}\}$, we can construct $F'$ as follows:
All clauses of $(c \setminus \{SET_{L}) \cup \{l\}$ must be removed from $F$ and must be replaced with $S_x \cup S_{\bar{x}}$ that are obtained using the matching-to-clauses construction method. The number of removed clauses is $|SET_{L}|\times|SET_{C}|$, while the number of added clauses is $|SET_{L}| + |SET_{C}|$ proving (2). Since the BVA is the complement of BVE, by applying BVE on $x$ in $F'$, it re-produces (reverse) $F$, and BVE preserves logical equivalence proving (1). $\blacksquare$

One of the best fitting applications of the BVA algorithm is re-encoding cardinality constraints \cite{aloul2002solving, chai2005fast, manthey2012automated, biere2014detecting}, where it is necessary to encode numerical bounds ($\leq k(x_1, x_2, ..., x_p)$). As discussed previously, the one-layer linear encoding formulates each keyRB output as the most special case of cardinality constraints ($k=1$), called \emph{at-most-1 constraint}. Compared to naive encoding for \emph{at-most-1 constraint} in which the number of clauses is $n.(n+1)/2$, by using BVA algorithm, the number of clauses would be reduced to $\sim3n$ \cite{manthey2012automated}. 

It is worth mentioning that numerous studies are explaining how cardinality constraints (numerical bounds) could be encoded efficiently \cite{bailleux2003efficient, sinz2005towards, marques2007towards, asin2011cardinality, koshimura2012qmaxsat}. Also, a few SAT solvers handle cardinality constraints by itself, such as Sat4J \cite{le2010sat4j} or clasp \cite{gebser2009conflict}; however, since these solvers do not extract cardinality constraints from the formula, compared to the direct re-encoding using BVA, their efficiency is extremely low. Furthermore, the strongest SAT solvers tend to not support native cardinality constraints, such as MiniSAT \cite{een2003extensible} that was supporting native cardinality constraints up to version 1.12. In \emph{CP\&SAT attack}, we employ the simpleBVA proposed in \cite{manthey2012automated} as a pre-processing step before running the SAT attack and after one-layer linear encoding. The simpleBVA will be used for each sub-CNF, corresponded to each keyRB output, and encoded using one-layer linear encoding, separately.

\subsubsection{SAT Execution and Key Matching}
\textcolor{white}{\vspace{8pt} \\}
After the reduction using the BVA algorithm, we update the CNF of the whole circuit using the reduced sub-CNFs corresponded to keyRB(s) outputs. Now, it is time to run the traditional SAT attack on the updated CNF. Since each keyRB might add cycles into the design, as mentioned in Section \ref{threat}, we need to use a cyclic-based SAT attack. Assuming that the circuits are acyclic, the CycSAT-I\footnote{CycSAT-I is designed and applicable to cyclic obfuscation when the original circuit is acyclic.} will be used.

As was mentioned previously, in one-layer linear encoding, the actual keys of each keyRB will be replaced with a set of one-hot key controlling the MUXes (one-layer encoding). Hence, after de-obfuscating the updated CNF, the SAT attack will recover the values of the one-hot keys. These one-hot keys determine the correct wiring/interconnection for the MUXes. So, a matching step is required, in which we need to calculate the actual key for each keyRB that establishes the same (correct) wiring/interconnection built by one-hot key in MUXes.  

\section{InterLock: Resisting CP\&SAT Attack} \label{proposed_defense}

In the previous section, we described how prior routing-based obfuscation techniques could be broken using the proposed \emph{CP\&SAT}. It calls into a question that \emph{"How routing obfuscation could be still used while it is not vulnerable to the BVA algorithm?"}. In this section, we answer this question by proposing a countermeasure that improves the resiliency of this category of obfuscation techniques against \emph{CP\&SAT} attack.

\subsection{Truly-Twisted Logic \& Routing Obfuscation}

To still get the benefit of routing obfuscation, and to combat against the efficiency of BVA-based re-encoding on routing-based obfuscation, we truly twist the keyRB with logic gates, meaning that a part of the actual logic gates will be also embedded into the keyRB. 

\begin{figure}[b]
    \centering
    \vspace{-20pt}
    \subfloat[key-programmable Routing Block (KeyRB) in Full-Lock \cite{kamali2019full}]{{\includegraphics[width=\columnwidth]{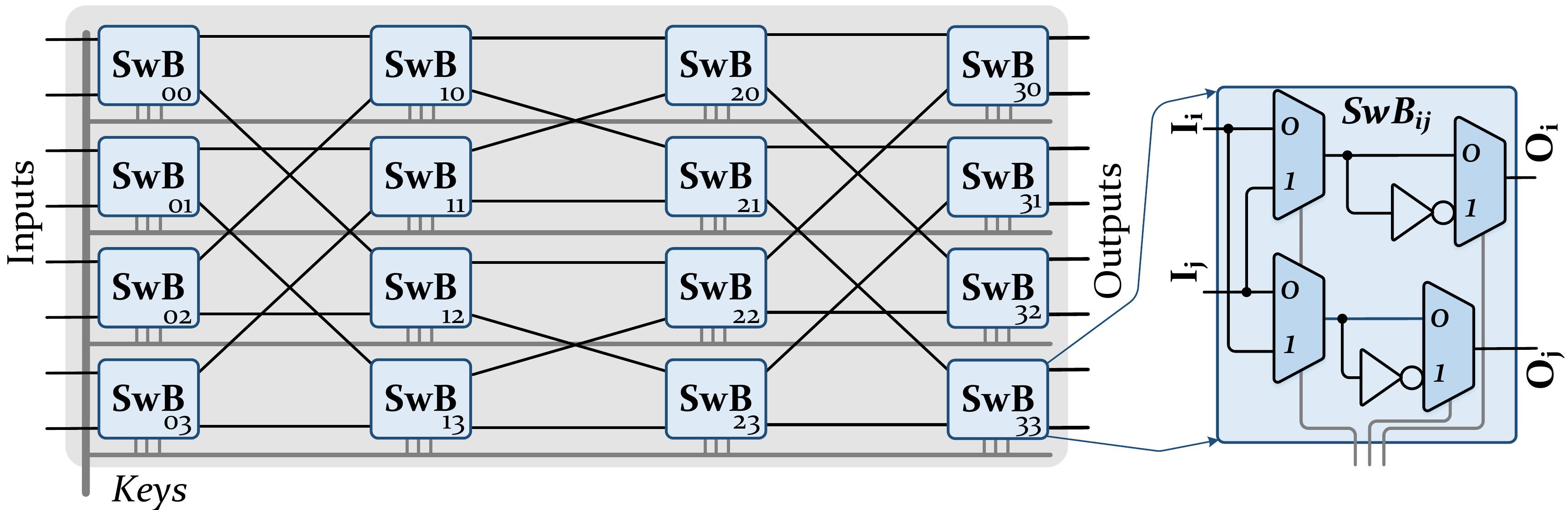}}} \\
    \subfloat[key-programmable Routing Block (KeyRB) in our Proposed InterLock]{{\includegraphics[width=\columnwidth]{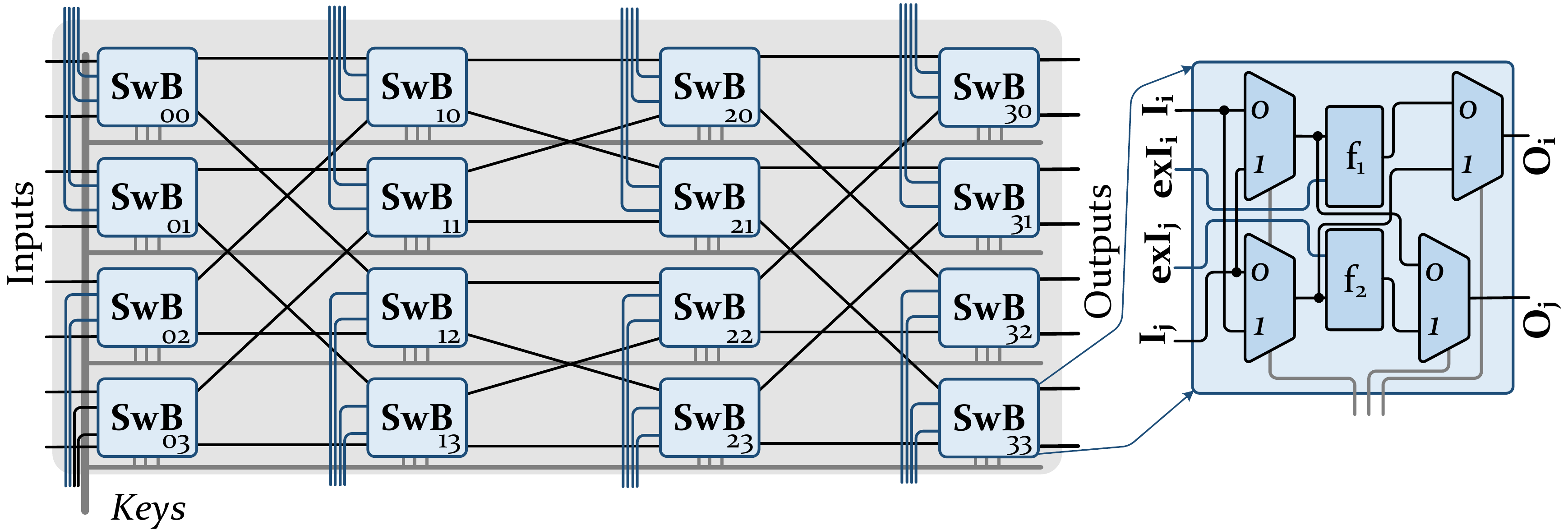}}}
    \caption{Full-Lock \cite{kamali2019full} vs. Our Proposed InterLock.}
    \label{swb_new}
\end{figure}

In the \emph{CP\&SAT} attack, we explained how a keyRB could be modeled as multiple numerical bound problems before the BVA re-encoding. So, the idea is that when the routing-based obfuscated circuit could not be translated (converted) to a numerical bound problem, the BVA is no longer applicable to it. For this purpose, inspired by the logarithmic (permutation) networks proposed in Full-Lock \cite{kamali2019full}, we employ the same architecture for keyRBs in InterLock; however, for each layer of that hierarchy, we will add a Boolean function (logic gate). Fig. \ref{swb_new} shows our new keyRB architecture that must be used for routing-based obfuscation. Compared to Full-Lock \cite{kamali2019full}, for each switch-box (SwB), the configurable inverters are removed, and $f_1$ and $f_2$ are added that could be any of 2-input basic logic gates, i.e. \emph{NAND, NOR, XNOR, AND, OR, XOR}. Also, for each SwB, we add extra inputs ($exI$) as one of the inputs of 2-input logic gates. For each SwB with 4 inputs ($I_i, I_j, exI_i, exI_j$), output $O_i$ could be \{$I_i$, $I_j$, $f_1(I_i, exI_i)$, and $f_1(I_j, exI_i)$\}, and output $O_j$ could be \{$I_i$, $I_j$, $f_2(I_i, exI_j)$, and $f_2(I_j, exI_j)$\}.

\subsubsection{Different Possibilities for $f_1$ and $f_2$}
\textcolor{white}{\vspace{8pt} \\}
In this section, we aim to explain (1) \emph{"Why the usage of $f_1$ and $f_2$ gates in the SwBs improves the resiliency of Interlock (compared to full-lock in which only routing and inversion is implemented)?"}, and (2) \emph{"how the selection of the logic for $f_1$ and $f_2$ affects its resiliency"}. To answer these two questions, we investigate five different scenarios: $f_1$s and $f_2$s could be (1) still inverters with no extra input (similar to Full-Lock), (2) all \emph{NAND} (\emph{AND}), (3)  all \emph{NOR} (\emph{OR}), (4)  all \emph{XNOR} (\emph{XOR}), and (5) selected randomly (any arbitrary 2-input gate). 

Table \ref{gate_explore} shows the runtime of the SAT attack (CycSAT-I) on an obfuscated ISCAS-85 c7552 circuit while only one keyRB is embedded into the design based on these five scenarios. The first and the most promising observation is that, compared to Full-Lock (when the logic layer is still inverter), for the same-size keyRB, the InterLock (all scenarios) builds a much harder SAT problem. As shown, in Full-Lock, the smallest single keyRB that breaks the SAT attack is a $64\times64$ keyRB (keyRB-64). However, in InterLock, it is even smaller (keyRB-16 or keyRB-32). Furthermore, we observe that, while all $f_1$s and $f_2$s are \emph{XNOR} (or \emph{XOR}), keyRB-16 is enough; however, for other gate types (\emph{NAND, NOR, AND, OR}), the smallest resilient is keyRB-32.

\begin{table}[t]
\footnotesize
\centering
\caption{The SAT Attack Runtime on ISCAS-85 c7552 with only One KeyRB in Different Scenarios.}
\label{gate_explore}
\setlength\tabcolsep{7pt} % default value: 6pt
\begin{tabular}{@{} *6c @{}}
\toprule
\hspace{0.5pt} \multirow{2}{*}{\backslashbox[30pt]{Size}{$f_{1,2}$}} & Full-Lock & InterLock  & InterLock & InterLock & InterLock \\
 & \cite{kamali2019full}  & all \emph{NAND} & all \emph{NOR} & all \emph{XNOR} & Random\\
\cmidrule(lr){1-1} \cmidrule(lr){2-6}
keyRB-4 & 0.02 & 0.192 & 0.136 & 0.718 & 0.232 \\ 
\cmidrule(lr){1-1} 
keyRB-8 & 0.437 & 3.083 & 5.905 & 2062 & 19.79 \\
\cmidrule(lr){1-1} \cmidrule(lr){5-5}
keyRB-16 & 5.413 & 522.1 & 558.2 & \textcolor{red}{Timeout} & 62332 \\
\cmidrule(lr){1-1} \cmidrule(lr){3-3} \cmidrule(lr){4-4} \cmidrule(lr){6-6}
keyRB-32 & 195.1 & \textcolor{red}{Timeout} & \textcolor{red}{Timeout} & \textcolor{red}{Timeout} & \textcolor{red}{Timeout} \\
\cmidrule(lr){1-1}  \cmidrule(lr){2-2}
keyRB-64 & \textcolor{red}{Timeout} & \textcolor{red}{Timeout} & \textcolor{red}{Timeout} & \textcolor{red}{Timeout} & \textcolor{red}{Timeout} \\
\bottomrule
\multicolumn{5}{l}{\textit{\textcolor{red}{Timeout} =  $10^5$ Seconds $\approx$ 1 day}}
\end{tabular}
\vspace{-10pt}
\end{table}

\subsubsection{Embedding Actual Timing Paths into KeyRBs}
\textcolor{white}{\vspace{8pt} \\}
This is a key observation that when all $f_1$s and $f_2$s are \emph{XNOR} (\emph{XOR}), the SAT resiliency of the keyRB is extremely higher. But, as shown in Fig. \ref{swb_new}, similar to Full-lock that engaged inverters to handle the toggling of some gates preceding the keyRB, in InterLock, these extra gates must become a part of the actual logic gates to avoid far exceeding the overhead. However, it is less likely to find a set of paths that only consist of \emph{XNOR} (\emph{XOR}). Hence, if we select the gate of each SwB based on an actual gate in a selected timing path, all $f_1$s and $f_2$s will become the actual gates of the design. It guarantees that, in InterLock, only MUXes could be considered as the overhead, however, in Full-Lock, all inversions except one layer are surplus as an extra overhead. Hence, although InterLock adds extra logic, it even reduces the overhead compared to the Full-Lock.

\begin{figure}
    \centering
    \subfloat[A Timing Path Selected to be Embedded into KeyRB]{{\includegraphics[width=\columnwidth]{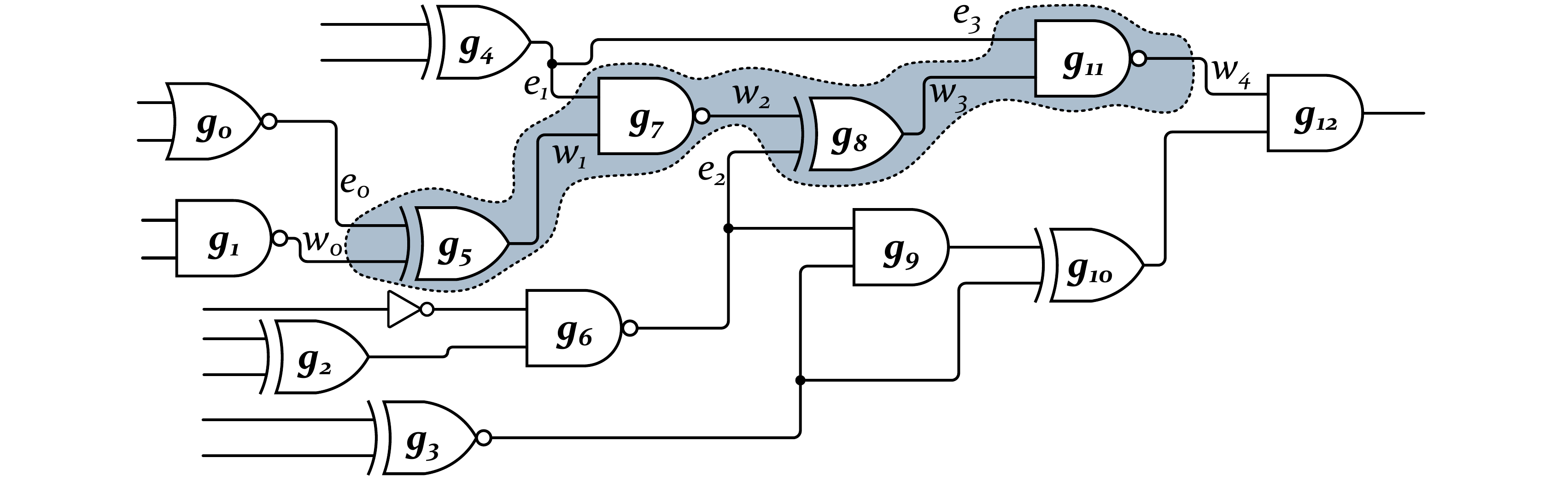}}} \vspace{-7pt}\\
    \subfloat[Inserting the Selected Timing Path into a KeyRB]{{\includegraphics[width=\columnwidth]{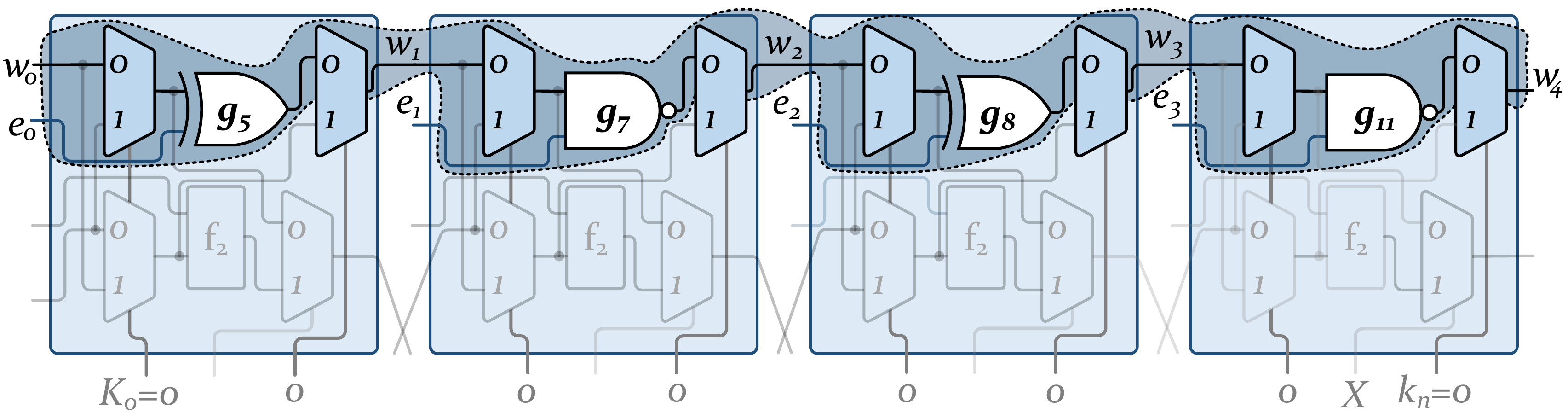}}} \vspace{-7pt}\\
    \subfloat[100\% Utilization: $m$ Timing Paths in a $m \times m$ KeyRB]{{\includegraphics[width=\columnwidth]{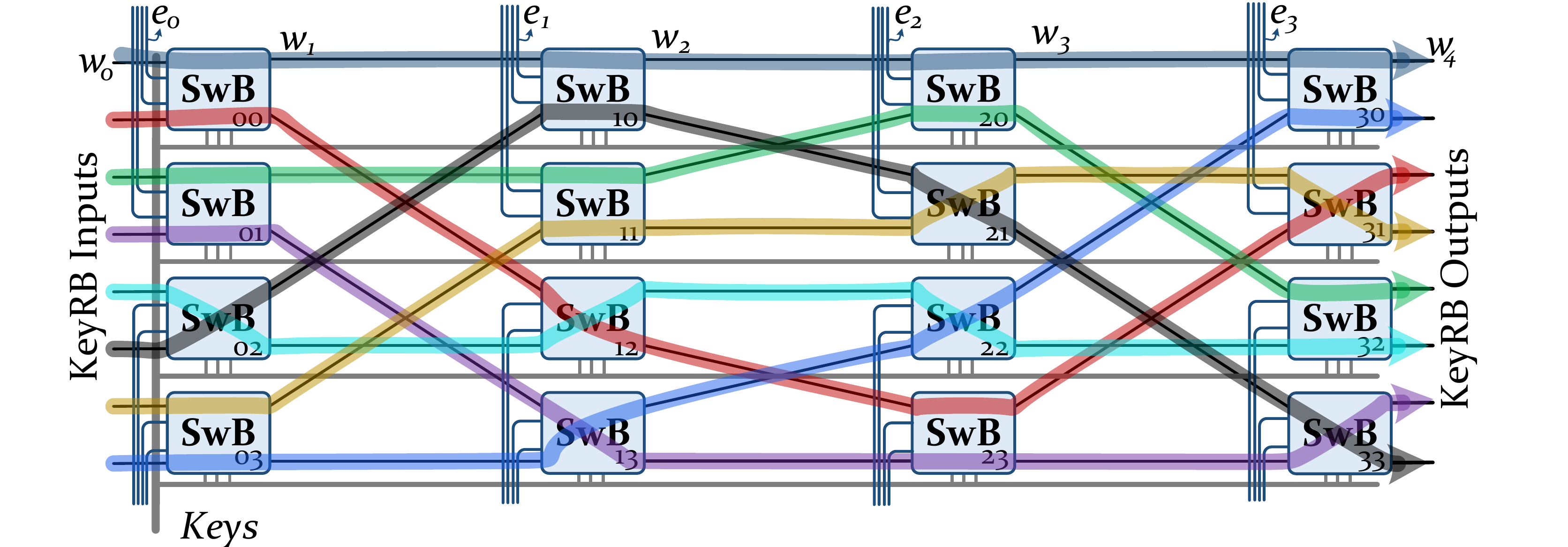}}} \vspace{-7pt}\\
    \caption{Timing Path Embedding into KeyRB.}
    \vspace{-15pt}
    \label{timing_path}
\end{figure}

To embed part(s) of the logic gates into each keyRB, we need a strategy to select the gates from the design. For the selection strategy, when the number of timing paths in each keyRB is $m$, $m$ timing paths must be selected\footnote{For a permutation-based $ m\times m$ network, there are $m$ different timing paths.} For $2log_2(m)-2$ layers of SwBs in permutation-based network \cite{kamali2019full}\footnote{The number of layers depends on the topology and being a blocking/non-blocking network. In this work, we use $2log_2(m)-2$ layers of SwBs for $m\times m$ network \cite{kamali2019full} that builds a near non-blocking permutation network.}, the length of the timing path must be equal with $2log_2(m)-2$. Hence, among the candidate timing paths, we select paths (or cut the paths) with length $2log_2(m)-2$. For example, Fig. \ref{timing_path} shows how an actual timing path will be embedded into a keyRB in InterLock. For a $8 \times 8$ keyRB, we have $2log_2(m)-2 = 2(3) - 2 = 4$ layers of SwBs. So, the timing paths must be the length of 4, and Fig. \ref{timing_path}(a) shows a part of the timing path with a length of 4 that is selected to be embedded into the keyRB. Fig. \ref{timing_path}(b) shows how this timing path is embedded into the keyRB. By using this approach, we embed $m$ timing paths into a $m\times m$ keyRB allowing us to utilize the logic gates of each keyRB by 100\%. Fig. \ref{timing_path}(c) shows the top view of 8 different paths that are embedded into a keyRB while an arbitrary key has determined the connection between keyRB I/O. 

\subsubsection{Twisted Logic in Interlock vs. Full-Lock}
\textcolor{white}{\vspace{8pt} \\}
In Full-Lock \cite{kamali2019full}, it is claimed that by adding a layer of configurable inverters into each SwB, the logic could be twisted with the keyRB. For example, Fig. \ref{full_lock_example} shows that gates $g_1$ and $g_6$ are converted to its negated model (\{\emph{AND, NOT}\} $\rightarrow$ \{\emph{NAND, BUFF}\}, and the inversion is handled within keyRB. Hence, to handle the inversion inside each keyRB, a layer of inversion is added into each SwB. However, such usage of the sequence of key-programmable inversion does not elevate the security of Full-Lock, and the KeyRB of Full-Lock could be simplified. More precisely, to attack the Full-lock, one could remove all inverters from the KeyRB (from all layers), and just add one layer of the key-programmable inverters at the end to reduce the key size while still maintaining the same function. This decouples the inversion from the routing block. Hence, in Full-Lock, the logic (inversion) and routing are not truly twisted. This allows us to simplify the KeyRB of Full-Lock before applying our \emph{CP\&SAT} attack to give maximum efficiency to the BVA. However, InterLock does not allow such simplification as $f_1$ and $f_2$ functions in each SwB are 2-input logic gates, and each input is (or could be considered as a) random and independent input.

\subsection{Area/Delay Overhead Exploration}

\begin{figure}
    \centering
    \subfloat[]{{\includegraphics[width=0.155\columnwidth]{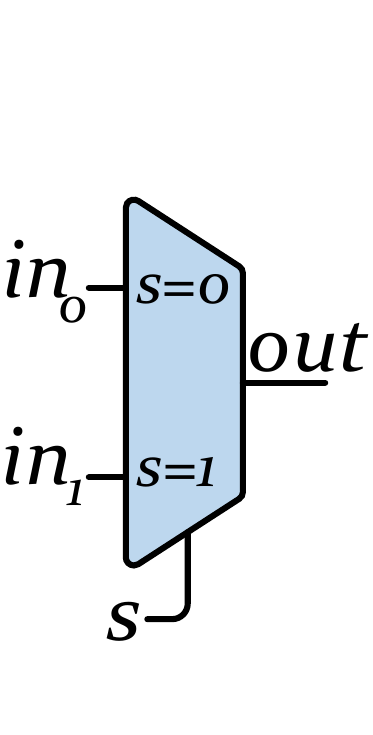}}} 
    \subfloat[]{{\includegraphics[width=0.165\columnwidth]{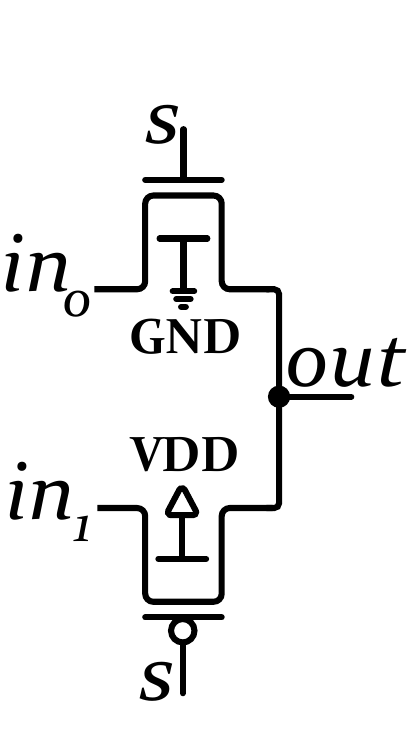}}} 
    \subfloat[]{{\includegraphics[width=0.175\columnwidth]{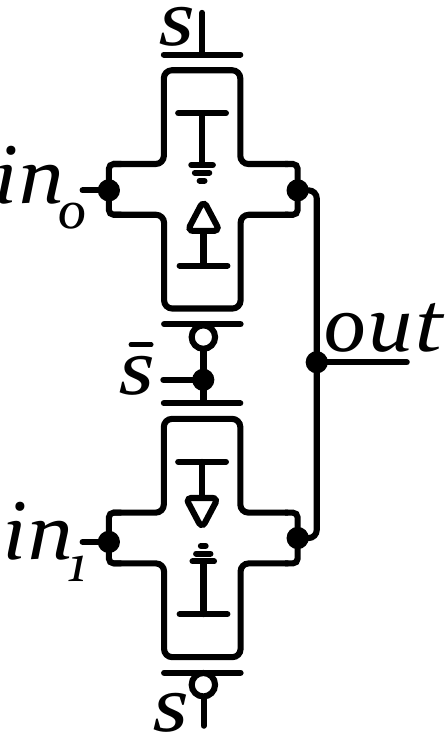}}} 
    \subfloat[]{{\includegraphics[width=0.195\columnwidth]{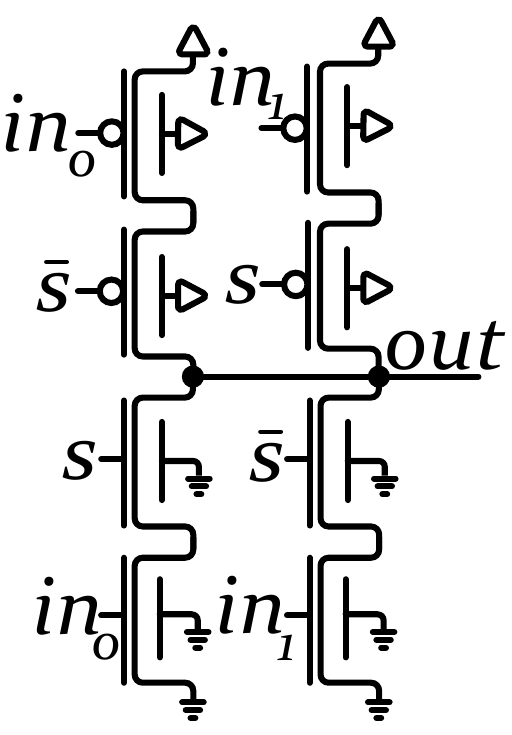}}} 
    \subfloat[]{{\includegraphics[width=0.125\columnwidth]{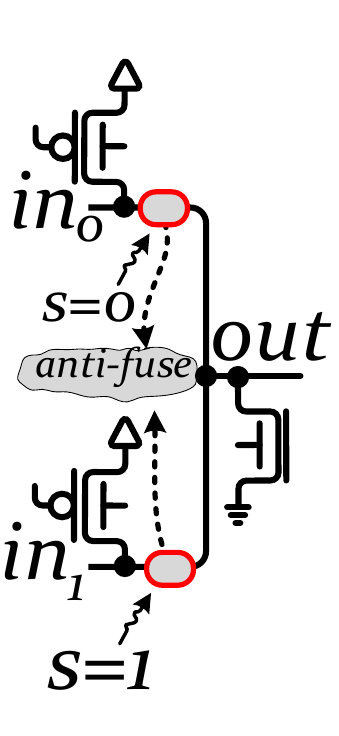}}} 
    \subfloat[]{{\includegraphics[width=0.185\columnwidth]{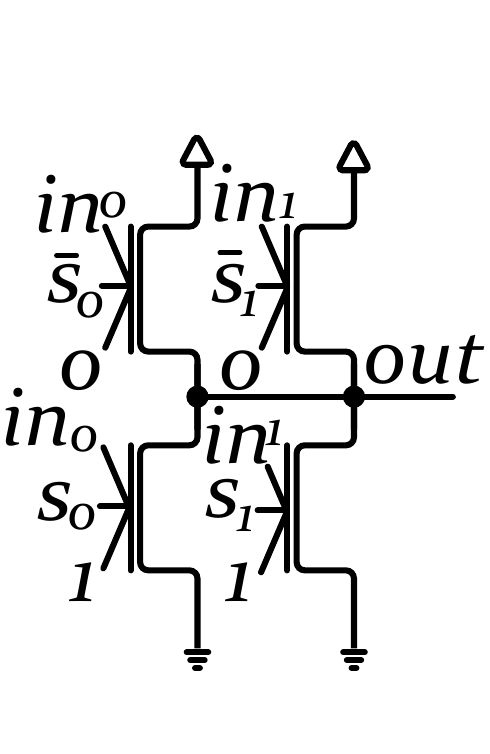}}} \vspace{-7pt}
    \caption{Different Multiplexer Implementation Possibilities: (a) 2:1 MUX Symbol, (b) 2:1 Pass-Transistor CMOS MUX, (c) 2:1 Transmission-Gate CMOS MUX, (d) 2:1 Static CMOS MUX, (e) 2:1 Anti-fused MUX, and (f) 2:1 TIGFET MUX. \vspace{-15pt}}
    \label{muxes}
\end{figure}

At first glance, embedding routing-based obfuscation incurs prohibited area/delay overhead. However, both Full-Lock and Cross-Lock engages some techniques to reduce the overhead to a reasonable ratio. Full-Lock shows how LUT insertion succeeding each keyRB allows them to use a smaller size of the keyRB to guarantee the resiliency at lower overhead. Unlike Full-lock that is implemented at gate-level and based on static CMOS technology, Cross-Lock engages anti-fuse-based elements called programmable via (PVIA) elements to minimize the overhead ratio of each keyRB. To implement InterLock in this paper, we examine both CMOS-based and PVIA-based implementation of keyRBs. Since MUXes are the only gate types that are used (overhead) for InterLock implementation, for CMOS-based implementation, amongst static logic, pass-transistor, or transmission gates, as demonstrated in Fig. \ref{muxes}(a)-\ref{muxes}(d), we engage transmission-gates (Tgate) for MUXes based on tree-like structure \cite{lee2008interconnect, luu2014vtr} that incur much lower overhead compared to static CMOS implementation. Also, as shown in Fig. \ref{muxes}(e), by using one-time-programmable elements (called PVIA elements in Cross-Lock \cite{shamsi2018cross}), we investigate the overhead of InterLock while implemented using anti-fuse-based (PVIA-based) 2:1 MUX. Similar to Tgate CMOS technology, we would use the tree-like structure to build keyRBs using PVIA elements.

Apart from these two technologies, in this paper, we assess the efficiency of another technology, called \emph{Three-Independent-Gate Field Effect Transistors (TIGFET)}, for implementing MUXes in InterLock. In TIGFET technology, each transistor has \emph{three} independent gates, and any two CMOS transistors could be modeled using only one TIGFET transistors, compacting the structure and achieving area as well as energy reduction, particularly for MUXes. Fig. \ref{muxes}(f) shows a 2:1 TIGFET multiplexer, and comparing with static CMOS each driving path consists of only one TIGFET transistors.

\begin{figure}[b]
    \centering
    \vspace{-25pt}
    \subfloat[]{{\includegraphics[width=0.15\columnwidth]{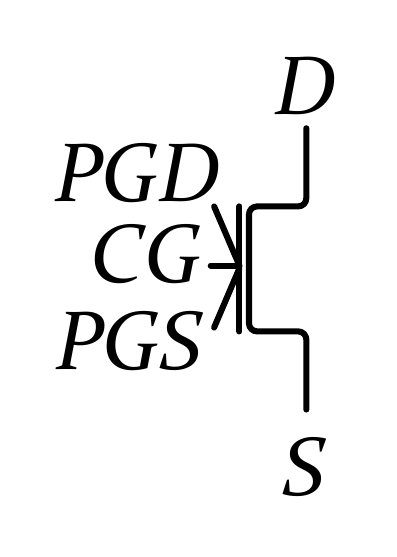}}}
    \subfloat[]{{\includegraphics[width=0.23\columnwidth]{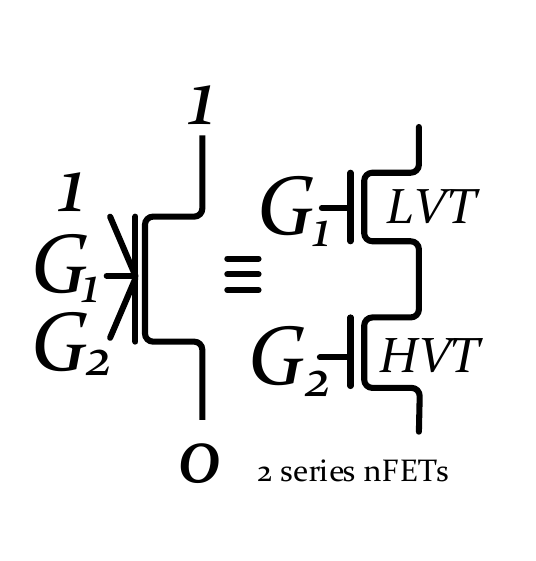}}}
    \subfloat[]{{\includegraphics[width=0.24\columnwidth]{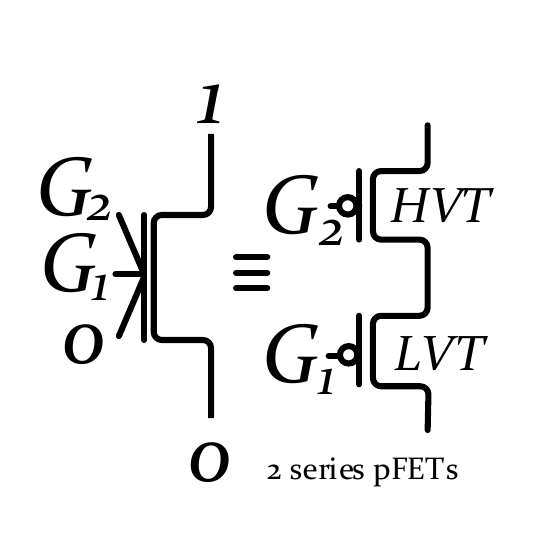}}}
    \subfloat[]{{\includegraphics[width=0.185\columnwidth]{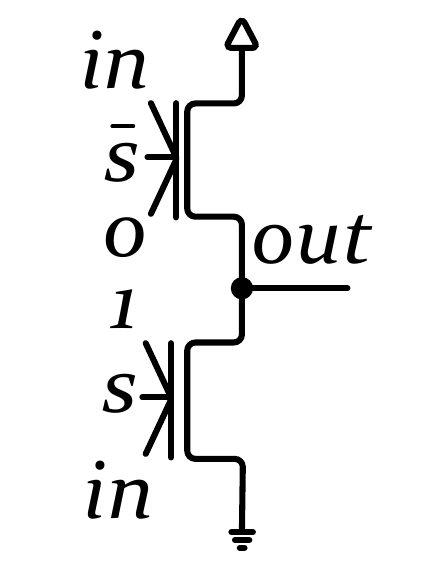}}}
    \subfloat[]{{\includegraphics[width=0.22\columnwidth]{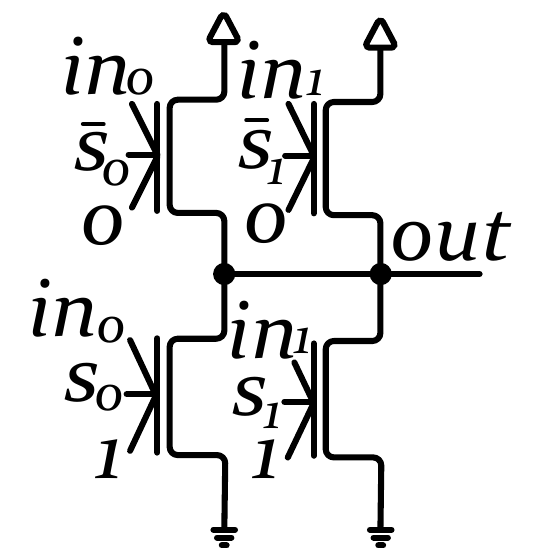}}}  \vspace{-7pt}
    \caption{2:1 TIGFET MUX Implementation: (a) A TIGFET Transistor, (b) 2 Series nFETs with One TIGFET Transistor, (c) 2 Series pFETs with One TIGFET Transistor, (d) A Tristate Inverter built by TIGFET Transistors, (e) 2:1 TIGFET MUX by Cascading TIGFET Tristate Inverters.}
    \label{tigfet}
\end{figure}

As shown in Fig. \ref{tigfet}(a), Three terminal gates in TIGFET transistors called \emph{Control Gate} (CG), \emph{Polarity Gate at Drain} (PGD), and \emph{Polarity Gate at Source} (PGS). Based on the value of these terminals, as illustrated in Fig. \ref{tigfet}(b, c), one could build 2 series nFETs or 2-series pFETs. Since MUXes could be built using tristate inverters, in Fig. \ref{tigfet}(d), we show how a tristate inverter could be built using TIGFET transistors. Compared to CMOS-based tristate inverter, the number of transistors is reduced by 50\% in the TIGFET version. When $m$ tristate TIGFET inverters are cascaded, a $m:1$ MUX is built (e.g. 2:1 TIGFET MUX in Fig. \ref{tigfet}(e)). It is worth mentioning that since the tristate inverter is used for each path of the multiplexer, the control signal (MUX selector) needs to be decoded. Hence, since all MUXes are controlled by the keys, and since we only use 2:1 MUXes, decoding selectors doubles the number of selectors (key inputs) in this technology. In our experimental results, we compare the implementation and the overhead of all three technologies.

\section{Experimental Results} \label{experiment}

To evaluate our proposed \emph{CP\&SAT} attack, we engage well-known ISCAS-89 and ITC-99 combinational circuits locked using Full-Lock \cite{kamali2019full}\footnote{Since Cross-Lock is a weaker version of Full-Lock (It has no configurable inverters), we only report the attack results on circuits locked by Full-Lock.}\footnote{Since all ISCAS-89 and ITC-99 are sequential, we apply all techniques on combinational parts of these circuits, and we assume that all FFs are accesible to be read/written for both SAT and \emph{CP\&SAT} attack.}. We sweep the size of keyRBs to show the efficiency of the BVA algorithm on routing-based obfuscation. Further, for our proposed \emph{InterLock}, as a countermeasure, we implement keyRBs from Fig. \ref{swb_new}(b) on the same benchmark circuits to acquire locked circuits. We apply both the SAT (CycSAT-I) and our proposed \emph{CP\&SAT} attack on locked circuits by InterLock. All the experiments are implemented using Python/C++ and have been carried out on many Dell PowerEdge R620 equipped with Intel Xeon E5-2670 2.50GHz and 64GB of RAM. We evaluate the overhead of InterLock in three different technologies: (1) Transmission-Gate (Tgate) CMOS using Synopsys generic 32nm library, (2) PVIA-based MUXes that are manually added between the M2 and M3 layers as physical-only cells, and (3) Silicon NanoWire TIGFETs (TIG SiNWFETs 32nm) modeled using Verilog-A \cite{zhang2014configurable, giacomin2017low}. 

\subsection{The Efficiency of the BVA}

To show how the BVA algorithm efficiently reduces the CNF size of the routing-based obfuscated circuits, Table \ref{cnf_reduced} shows the rate of reduction that is more than a factor of two. The BVA adds/substitute the variables to decrease the number of clauses. As shown, the number of variables increases (by up to 2x); however, the number of clauses that play an important role in determining the complexity of the CNF is decreased by more than 2x. Hence, the BVA-based pre-processed CNFs are far easier for the SAT solver to be solved. Furthermore, as can be seen in Table \ref{cnf_reduced}, increasing the size of the keyRBs does not increase the SAT runtime after BVA pre-processing exponentially (most likely quadratically). Since we cannot infinitely increase the size of the keyRB (due to overhead and limitation of candidate selection), we report the results on keyRB with size up to 64$\times$64. 

\begin{table}[h]
\footnotesize
\centering
\caption{The Effectiveness of the BVA Pre-Processing Step on Different-Size keyRBs in Routing-based Obfuscation.\vspace{-7pt}}
\label{cnf_reduced}
\setlength\tabcolsep{3pt} % default value: 6pt
\begin{tabular}{@{} l cccccc @{}}
\toprule
\hspace{0.5pt} \multirow{2}{*}{Instance} & \multicolumn{3}{c}{Original} & \multicolumn{3}{c}{BVA Pre-processed} \\
\cmidrule(lr){2-4} \cmidrule(lr){5-7}
& \#Variables & \#Clauses & Solve & \#Variables & \#Clauses & pre+Solve \\
\cmidrule(lr){1-1} \cmidrule(lr){2-4} \cmidrule(lr){5-7}
keyRB-4 & 271 & 418 & \textbf{0.02} & 428 & 202 & \textbf{0.01+0.22} \\
keyRB-8 & 875 & 1606 & \textbf{0.45} & 1278 & 718 & \textbf{0.01+0.36} \\
keyRB-12 & 1544 & 3084 & \textbf{2.48} & 2188 & 1288 & \textbf{0.01+0.54} \\
keyRB-16 & 2419 & 4750 & \textbf{5.42} & 3982 & 2184 & \textbf{0.01+0.82} \\
keyRB-24 & 3372 & 7502 & \textbf{54.82} & 4618 & 3452 & \textbf{0.02+1.64} \\
keyRB-32 & 6178 & 12510 & \textbf{194.8} & 8892 & 7258 & \textbf{0.02+2.22} \\
keyRB-48 & 9891 & 18614 & \textbf{Timeout} & 12672 & 9918 & \textbf{0.04+3.92} \\
keyRB-64 & 15043 & 31182 & \textbf{Timeout} & 23818 & 14772 & \textbf{0.04+12.22} \\
\bottomrule
\multicolumn{5}{l}{\textit{Timeout =  $10^5$ Seconds $\approx$ 1 day}}
\end{tabular}
\end{table}

To show the success of our proposed \emph{CP\&SAT} attack on routing-based obfuscation, in Table \ref{canonical_runtime}, we illustrate the runtime of the SAT (CycSAT-I) and the \emph{CP\&SAT} attack on circuits locked by Full-Lock. As shown, in all cases, after inserting four keyRB-16 (16$\times$16), the traditional SAT attack fails to break the locked circuits. However, when we apply the \emph{CP\&SAT} attack, all circuits locked by four keyRB-16 are broken in less than 10 minutes. When we assume that the configurable inverters of SwBs in Full-Lock are in place, the BVA algorithm within the \emph{CP\&SAT} attack does not provide a significant advantage. As shown in Table \ref{canonical_runtime}, we observed that this new attack also fails to break circuits locked with four keyRB-16 while inverters are intact. However, as described previously, we detach the inverters in Full-Lock by fixating the key values of all layers of inverters (disable the inversion) except the last layer. When the inverters are detached, the BVA could efficiently reduce the size of the locked circuit allowing us to de-obfuscate within few minutes.

\begin{table}[h]
\footnotesize
\centering
\caption{The Runtime of the SAT Attack and the \emph{Cnonical Prune-and-SAT} Attack on Circuits Locked by Full-Lock \cite{kamali2019full} with Different Sizes of KeyRBs. \vspace{-7pt}}
\label{canonical_runtime}
\setlength\tabcolsep{1.3pt} % default value: 6pt
\begin{tabular}{@{} ccc ccc ccc ccc @{}}
\toprule
\hspace{0.5pt} \multirow{5}{*}{Circuit} & \multirow{5}{*}{\#Gates} & \multirow{5}{*}{\#I/O} & \multicolumn{3}{c}{Traditional SAT} & \multicolumn{3}{c}{proposed \emph{canonical}} & \multicolumn{3}{c}{proposed \emph{canonical}} \\
& & & \multicolumn{3}{c}{Attack} &  \multicolumn{3}{c}{\emph{prune\&SAT} } &  \multicolumn{3}{c}{\emph{prune\&SAT} } \\
& & & \multicolumn{3}{c}{(CycSAT-I)} & \multicolumn{3}{c}{(with inverters)} & \multicolumn{3}{c}{(detached inverters)} \\
\cmidrule(lr){4-6} \cmidrule(lr){7-9} \cmidrule(lr){10-12}
& & & \multicolumn{3}{c}{keyRB-16 (16$\times$16)} & \multicolumn{3}{c}{keyRB-16} & \multicolumn{3}{c}{keyRB-16} \\
\cmidrule(lr){4-6} \cmidrule(lr){7-9} \cmidrule(lr){10-12}
& & & 2 & 3 & 4 & 2 & 3 & 4 & 2 & 3 & 4 \\
\cmidrule(lr){1-1} \cmidrule(lr){2-3} \cmidrule(lr){4-6} \cmidrule(lr){7-9} \cmidrule(lr){10-12}

b15 & $\sim$8.5K & 485/519 & 1507.5 & TO & TO & 486.4 & 2581.5 & TO & 44.8 & 75.9 & 319.8 \\

b14 & $\sim$9.5K & 277/299 & 788.3 & TO & TO & 329.7 & 1688.8 & TO & 34.8 & 88.9 & 416.6 \\

s35932 & $\sim$16K & 1763/2048 & 856.6 & TO & TO & 643.8 & 5238.1 & TO & 76.4 & 147.9 & 407.4 \\

s38417 & $\sim$18K & 1464/1731 & 1187.4 & TO & TO & 482.5 & 2037.9 & TO & 58.2 & 100.7 & 366.3 \\

b20 & $\sim$19.5K & 522/512 & 1096.8 & TO & TO & 537.8 & 3507.9 & TO & 70.8 & 129.4 & 411.2 \\

b21 & $\sim$20K & 522/512 & 1832.4 & 13283 & TO & 984.8 & 8207.3 & TO & 134.4 & 207.8 & 550.1 \\

b17 & $\sim$30K & 1452/1512 & 508.2 & 8401.7 & TO & 306.8 & 6095.4 & TO & 81.7 & 137.4 & 463.8 \\

b22 & $\sim$30K & 767/757 & 924.6 & 6491.5 & TO & 508.2 & 5538.4 & TO & 68.5 & 99.1 & 390.5 \\

b18 & $\sim$110K & 3357/3343 & 1283.7 & 9208.1 & TO & 581.9 & 6327.8 & TO & 91.6 & 162.7 & 472.2 \\

\bottomrule
\multicolumn{5}{l}{\textit{Timeout (TO) =  $10^5$ Seconds $\approx$ 1 day}}
\end{tabular}
\vspace{-12pt}
\end{table}

\subsection{Disabling the BVA using InterLock}

To show how InterLock could be used as a countermeasure against the \emph{Canonical prune and SAT} attack, we insert different numbers of the keyRBs from Fig. \ref{swb_new}(b) into the benchmark circuits with different sizes. For the insertion of the keyRBs two strategies have been considered/applied in InterLock: (1) To minimize the performance degradation (delay overhead), the timing paths that are selected as the candidates to be embedded into keyRBs must be one of the highest positive slack timing paths, and (2) as shown in Table \ref{runtime_simple}, since having more \emph{XNOR}s (\emph{XOR}s) increase the resilience of the locked circuits considerably, amongst the candidates, we select those paths that have more \emph{XNOR}s (\emph{XOR}s).  

Table \ref{sat_canonical_runtime_interlock} shows the runtime of both the SAT and the \emph{canonical prune and SAT} attack on circuits locked by InterLock. In both cases, since the locked circuit is possibly cyclic, for the SAT solving part, CycSAT-I has been used. As shown, for almost all cases, after inserting only two keyRB-16, both attacks fail to break the locked circuit. Unlike previous routing-based obfuscation techniques, BVA does not provide any advantage as a pre-processing step showing that truly twisting logic into the keyRBs guarantees the resistance against this new attack. Furthermore, compared to Full-Lock and Cross-Lock, twisting logic into keyRB allows us to engage smaller sizes of keyRB to guarantee the resiliency (keyRB-32/16 $\rightarrow$ keyRB-16/8). Shrinking the size of the keyRB with guaranteed security in InterLock allows the designer to considerably reduce area and delay overhead. 

\begin{table}[t]
\footnotesize
\centering
\caption{The Runtime of the SAT Attack as well as the \emph{Cnonical Prune-and-SAT} Attack on Circuits Locked by InterLock with Different Sizes of KeyRBs. \vspace{-10pt}}
\label{sat_canonical_runtime_interlock}
\setlength\tabcolsep{3pt} % default value: 6pt
\begin{tabular}{@{} c ccc cc ccc cc @{}}
\toprule
\hspace{0.5pt} \multirow{4}{*}{Circuit} &  \multicolumn{5}{c}{Traditional SAT} & \multicolumn{5}{c}{proposed \emph{canonical}} \\
& \multicolumn{5}{c}{Attack (CycSAT-I)} &  \multicolumn{5}{c}{\emph{prune\&SAT} }  \\
\cmidrule(lr){2-6} \cmidrule(lr){7-11}
& \multicolumn{3}{c}{keyRB-8} & \multicolumn{2}{c}{keyRB-16} & \multicolumn{3}{c}{keyRB-8} & \multicolumn{2}{c}{keyRB-16} \\
\cmidrule(lr){2-4} \cmidrule(lr){5-6} \cmidrule(lr){7-9} \cmidrule(lr){10-11}
& 1 & 2 & 3 & 1 & 2 & 1 & 2 & 3 & 1 & 2 \\
\cmidrule(lr){1-1} \cmidrule(lr){2-4} \cmidrule(lr){5-6} \cmidrule(lr){7-9} \cmidrule(lr){10-11}

b15 & 98.5 & 3807.8 & TO & 74127.8 & TO & 85.3 & 3207.8 & TO & 69328.5 & TO \\

b14 & 120.8 & 4229.1 & TO & 67203.2 & TO & 105.1 & 4028.5 & TO & 64328.6 & TO \\

s35932 & 291.7 & 7126.4 & TO & 59372.1 & TO & 260.7 & 6992.1 & TO & 55221.4 & TO \\

s38584 & 267.9 & 7624.4 & TO & 71375.5 & TO & 233.8 & 7168.5 & TO & 63298.8 & TO \\

b20 & 284.4 & 11275.8 & TO & 58348.6 & TO & 186.7 & 8673.8 & TO & 55373.3 & TO \\

b21 & 738.4 & TO & TO & TO & TO & 672.5 & TO & TO & TO & TO \\

b17 & 320.4 & 6221.3 & TO & 77023.3 & TO & 271.9 & 5882.2 & TO & 74623.7 & TO \\

b22 & 376.4 & 5209.9 & TO & 51042.4 & TO & 126.7 & 3862.7 & TO & 37621.2 & TO \\

b18 & 701.9 & 32841.5 & TO & TO & TO & 630.3 & 30067.7 & TO & TO & TO \\

\bottomrule
\end{tabular}
\end{table}

To illustrate that InterLock elevate the complexity of locked circuit, in Table \ref{iteration_compare}, we compare the average number of iterations ($N$ in Eq. \ref{runtime_simple}) in Full-Lock with that of InterLock. Since we still get the benefit of routing-based obfuscation, the number ($M$ in Eq. \ref{runtime_recursive}) and the computational complexity ($T_{DPLL}^{Avg}$ in Eq. \ref{runtime_recursive}) of recursive calls in DPLL algorithm is still extremely high in InterLock. However, as illustrated in Table \ref{iteration_compare}, the average number of required iterations is increased by $\sim$3x-4x. This increase shows that $MN \times T_{DPLL}^{Avg}$ (from Eq. \ref{runtime_recursive}) is extremely higher in InterLock deepening the logic locking problem significantly.

\begin{table}[t]
\footnotesize
\centering
\caption{The Average Number of Iterations for De-Obfuscating Different Sizes of KeyRBs. (Numbers in parentheses show the Current Iteration at Timeout).\vspace{-10pt}}
\label{iteration_compare}
\setlength\tabcolsep{5pt} % default value: 6pt
\begin{tabular}{@{} *6c @{}}
\toprule
\hspace{0.5pt} \multirow{2}{*}{\backslashbox[50pt]{Model}{Insance}} & \multirow{2}{*}{keyRB-4} & \multirow{2}{*}{keyRB-8}  & \multirow{2}{*}{keyRB-16} & \multirow{2}{*}{keyRB-32} & \multirow{2}{*}{keyRB-64} \\ \\
\cmidrule(lr){1-1} \cmidrule(lr){2-6}
Full-Lock \cite{kamali2019full} & 3-5 & 4-6 & 8-10 & 10-12 & (5) Timeout\\ 
\cmidrule(lr){1-1}  \cmidrule(lr){2-6} 
InterLock & 8-12 & 16-22 & 30-33 & (8) Timeout & (9) Timeout\\
\bottomrule
\end{tabular}
\end{table}

\subsection{Elevated Security at Lower Overhead}

To perform a proof-of-concept physical design flow, we implement the keyRB in InterLock using three different 32nm technology: (1) Transmission-Gate (Tgate) CMOS, (2) PVIA-based MUXes, and (3) TIGFET SiNWFETs using Verilog-A. Table \ref{overhead_interlock} shows the area/power/delay overhead of locked circuits via three keyRB-8, which is resilient against the SAT and the \emph{cnonical prune-and-SAT} attack. Compared to Full-Lock which is a gate-level implementation of MUXes using static CMOS, in InterLock, Tgate-based implementation at transistor level would considerably reduce the area/delay overhead. 
In many cases it was expected to observe that PVIA-based MUXes could achieve the most optimum results; However, since there is no automatic flow in existing EDA tools for optimization of a large number of PVIA-based elements, the insertion has to be done manually. To do it manually, we inserted the PVIAs in a grid and push the standard cells away from this PVIA grid to successfully perform placement, and due to fine-granularity of MUXes in the circuit (small units and a large amount of usage), and since the number of PVIAs that must be used is a lot, in many cases we faced DRC violations leading us to use much lower utilization rate for them.

\begin{table}[t]
\footnotesize
\centering
\caption{The Overhead Comparison between Three Different Technology used for InterLock Implementation: Tgate, Anti-fuse, and TIGFET. \vspace{-10pt}}
\label{overhead_interlock}
\setlength\tabcolsep{1.5pt} % default value: 6pt
\begin{tabular}{@{} c ccc ccc ccc ccc @{}}
\toprule

\hspace{0.5pt} \multirow{5}{*}{Circuit} & \multicolumn{3}{c}{\multirow{3}{*}{Original}} & \multicolumn{9}{c}{3$\times$keyRB-8} \\
\cmidrule(lr){5-13}  
& & & & \multicolumn{3}{c}{Tgate CMOS} & \multicolumn{3}{c}{PVIA Anti-Fuse} & \multicolumn{3}{c}{TIGFET} \\
 \cmidrule(lr){2-4} \cmidrule(lr){5-7} \cmidrule(lr){8-10} \cmidrule(lr){11-13}
& a & p & d & a & p & d & a & p & d & a & p & d \\
& ($\mu$m$^2$) & ($\mu$W) & (ns) & \% & \% & \% & \% & \% & \% & \% & \% & \% \\
\cmidrule(lr){1-1} \cmidrule(lr){2-4} \cmidrule(lr){5-7} \cmidrule(lr){8-10} \cmidrule(lr){11-13}
b15 & 5292.7 & 327.6 & 1.23 & 34.5\% & 27.8\% & 12.9\% & 27.6\% & 21.9\% & 7.1\% & 24.5\% & 20.1\% & 6.4\% \\
b14 & 5707.9 & 423.9 & 1.55 & 22.6\% & 17.5\% & 14.6\% & 19.5\% & 16.2\% & 8.8\% & 18.6\% & 15.4 & 6.8\% \\
s35932 & 9283.1 &  729.8 & 1.68 & 30.7\% &	22.8\% &	10.9\% & 27.3\% & 20.7\% & 6.5\% & 24.8\% &	18.7\% & 5.1\% \\ s38584 & 11003.2 & 806.6 & 1.77 & 24.1\% &	19.1\% &	19.7\% &	22.4\% &	17.3\% &	10.7\% &	20.7\% &	16.7\% &	8.3\% \\
b20 & 11752.5 & 755.6 & 2.34 & 19.8\% &	15.5\% &	12.8\% &	17.6\% &	13.9\% &	7.1\% &	15.9\% &	13.4\% &	5.6\% \\
b21 & 13007.1 & 922.7 & 2.21 & 16.2\% &	12.7\% &	10.7\% &	14.3\% &	11.2\% &	5.7\% &	12.9\% &	10.4\% &	4.5\% \\
b17 & 15573.3 & 1245.1 & 3.58 & 8.9\% &	7.4\% &	7.6\% &	7.8\% &	6.4\% &	4.2\% &	7.2\% &	5.9\% &3.4\% \\
b22 & 16582.7 & 1319.5 & 3.18 & 6.4\% &	4.9\% &	8.5\% & 5.9\% &	4.1\% &	4.3\% &	4.8\% &	3.9\% &	3.5\% \\
b18 & 57626.9 & 4834.1 & 3.81 & 3.7\% &	2.1\% &	3.2\% & 3.5\% & 1.8\% &	1.9\% &	2.7\% &	1.4\% &	1.6\% \\
\bottomrule
\end{tabular}
\end{table}

Based on our evaluation of these three different technologies, as shown in Table \ref{overhead_interlock}, TIGFET-based keyRB could bring more efficiency compared to Tgate CMOS and PVIA-based implementation. As shown, on average, TIGFET could reduce the area/delay overhead by up to 20\%/56\% compared to Tgate-based CMOS keyRB. However, for two important reasons, in all three technologies, on average, the overhead is less than 10\%, which is completely acceptable: (1) The required number/size of keyRBs is less/smaller in InterLock, and (2) The actual timing paths selected for embedding are paths with highest positive slack time. 

\section{Conclusion} \label{conclusion}

In this paper, we proposed a new attack, \emph{canonical prune-and-SAT} (\emph{CP\&SAT}) attack, for breaking the state-of-the-art routing-based obfuscation solutions (e.g. Full-Lock \cite{kamali2019full} and Cross-Lock \cite{shamsi2018cross}). In this attack, we exploited a \emph{bounded variable addition} (BVA) pre-processing step (before the SAT attack) to reduce the size and complexity of the CNF representation of the key-programmable routing blocks (keyRBs) used for routing-based obfuscation. We demonstrated that by adding the BVA pre-processing step, our proposed attack reduces the size of the targeted CNF formula by a factor of two, enabling us to easily break both Full-Lock and Cross-Lock in a reasonable time. Further, as a countermeasure, we proposed our unified routing and logic obfuscation techniques, coined as \emph{InterLock}. In Interlock, in addition to hiding the wiring/interconnection, a part of the logic (gates) in the selected timing paths are also implemented in the keyRB. We illustrated that when the logic gates are twisted with keyRBs, the BVA could not provide any advantage as a pre-processing step. We demonstrated that, by using InterLock, the resilience against existing attacks as well as our new proposed \emph{CP\&SAT} attack would be guaranteed while the delay/area overhead remains acceptable. We further evaluated and depicted the result of implementing the Interlock using three different technologies: Transmission-gate CMOS, anti-fuse elements, and three-independent-gate field-effect transistors (TIGFET).

\section*{Acknowledgement} \label{Acknowledgement}

This research is supported by the Defense Advanced Research Projects Agency (DARPA-AFRL, \#FA8650-18-1-7819), National Science Foundation (NSF, \#1718434), and partly by Silicon Research Co. (SRC TaskID 2772.001).

\bibliographystyle{ACM-Reference-Format}
\bibliography{refs}

\end{document}